\documentclass[letterpaper, 10 pt, conference]{ieeeconf}  

\IEEEoverridecommandlockouts  
\usepackage[pdftex]{graphicx}
\DeclareGraphicsExtensions{.pdf,.jpeg,.png}

\usepackage{amsmath} 
\usepackage{amssymb}  

\usepackage{tabularx,booktabs,multirow}
\usepackage{array}
\usepackage{float}
\usepackage{makecell}
\usepackage{tablefootnote}
\usepackage{multicol}
\usepackage{booktabs}
\usepackage{threeparttable}
\usepackage{url}
\usepackage[font=small,skip=10pt]{caption}
\usepackage[dvipsnames]{xcolor}
\usepackage{amsmath}
\usepackage{xcolor}
\usepackage{algorithmic}
\usepackage{dirtytalk}
\usepackage[ruled,vlined]{algorithm2e}

\newcolumntype{P}[1]{>{\centering\arraybackslash}p{#1}}
\SetKwInOut{Input}{input}
\SetKwInOut{Output}{output}
\makeatletter
\let\NAT@parse\undefined
\makeatother
\usepackage{hyperref}
\hypersetup{colorlinks,allcolors=red}

\title{\LARGE \bf
Intuitive Tasks Planning Using Visuo-Tactile Perception for Human Robot Cooperation
}

\author{Sunny Katyara$^{1,2}$, Fanny Ficuciello$^{2}$,~\IEEEmembership{Senior Member,~IEEE}, Tao Teng$^{3}$, Fei Chen$^{4}$,~\IEEEmembership{Senior Member,~IEEE}, \\
Bruno Siciliano$^{2}$,~\IEEEmembership{Fellow,~IEEE},  Darwin G. Caldwell$^{1}$,~\IEEEmembership{Senior Member,~IEEE}, 
\thanks{This research is supported by projects ``LEARN-REAL'' funded by EU H2020 ERA-Net Chist-Era and ``HARMONY'' funded by EU H2020 R\&I under grant agreement 101017008. \textit{(Corresponding author: Sunny Katyara)} }
\thanks{$^{1}$ Department of Advanced Robotics, Istituto Italiano di Tecnologia, Via Morego 30, 16163, Genova, Italy (e-mail: {\tt\small name.surname@iit.it}).}
\thanks{$^{2}$ Department of Information Technology and Electrical Engineering and PRISMA Lab, University of Naples Federico II, Naples 80125, Italy ({e-mail: \tt\small name.surname@unina.it}).}
\thanks{$^{3}$ Universita Cattolica del Sacro Cuore, Via E.Parmense 84, Piacenza, 29122, Italy ({e-mail: \tt\small name.surname@unina.it}).}
\thanks{$^{4}$ Department of Mechanical and Automation Engineering, The Chinese University of Hong Kong, Hong Kong ({e-mail: \tt\small f.chen@ieee.org}).}
}

\begin{document}

\maketitle
\thispagestyle{empty}
\pagestyle{empty}

\begin{abstract}

Designing robotic tasks for co-manipulation necessitates to exploit not only proprioceptive but also exteroceptive information for improved safety and autonomy. Following such instinct, this research proposes to formulate intuitive robotic tasks following human viewpoint by incorporating visuo-tactile perception. The visual data using depth cameras surveils and determines the object dimensions and human intentions while the tactile sensing ensures to maintain the desired contact to avoid slippage. Experiment performed on robot platform with human assistance under industrial settings validates the performance and applicability of proposed intuitive task formulation.    

\end{abstract}

\section{INTRODUCTION}

In last two decades, the robots have gained sufficient social trust and are extensively participating with humans in performing certain tasks that require cognitive abilities of humans to be combined with precision and strength of robots \cite{c1}\cite{c2}. However, such a social trust on robots requires to endow them with multi-sensory information especially the visuo-tactile feedback to make instant decisions, detect obstacles, recognize human intervention and adapt to varying environment proactively \cite{c3}\cite{c4}. The visuo-tactile data thus enables the robot to understand the non-verbal cues to collaborate more intuitively from human perspective.

Exploiting visuo-tactile information in cluttered environment for human-robot joint carrying task, a framework is proposed in \cite{c5}. Under this approach, the robot tasks are defined using standard stack-of-task (SoT) formulation but without taking into account the human intuition and ergonomics. In the same way, a modified technique is presented in \cite{c6} for industrial assembly tasks. Wherein, the adaptive gains and homotopy are introduced in conjunction with visuo-tactile data for switching human-robot roles smoothly according to task requirements but however it does not consider progressive mutations in the agents’ (i.e, robot and human) behavior and environment. Hence, to the extent of our knowledge, there is no any intuitive task formulation framework available that considers the human ergonomics and task progress in planning hierarchical robot actions, thereby exploiting visuo-tactile perception for fine co-manipulation tasks. The results presented here are the part of our recent works in \cite{c7}\cite{c8}. 

\section{Research Methodology}

   \begin{figure}[t]
      \centering
      \includegraphics[width=8.5cm]{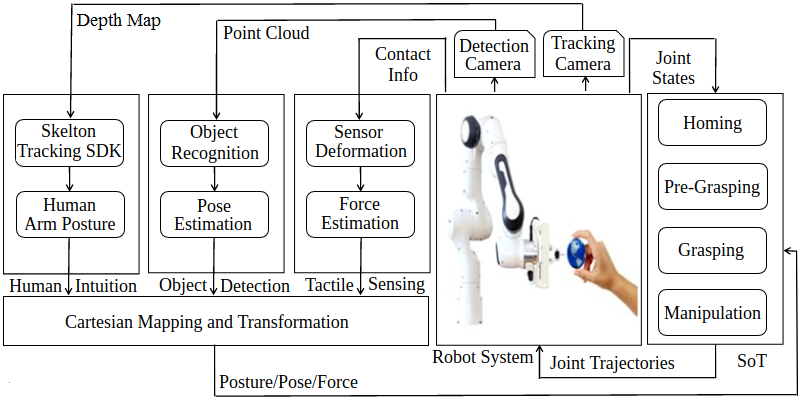}
      \caption{Proposed research methodology on formulating intuitive robot actions using an idea of SoT for co-manipulation using visual data for human intuition estimation and object detection and the tactile sensing for flexible object interaction. Human gestures are determined using a standard skeleton tracking algorithm. For object detection, RANSAC algorithm together with SVM classifier is used while the tactile sensing ensure to modulate the gripper's force profile to avoid object slippage based on the friction cone criteria.}
      \label{three}
      \vspace{-15pt}
   \end{figure}

The robot tasks are defined in two groups i.e, the Cartesian tasks accounting for position and orientation of end-effector and the force tasks ensuring flexible and adaptive interaction with the human and objects in the environment. All the primary and secondary tasks are defined in a stack with hard and soft priorities being assigned to each at different levels and are executed sequentially, following a standard hierarchical control formulation called SoT. However, the tasks in SoT framework are Quadratic Programming (QP) problems and are formulated according to \cite{c9}\cite{c10}.

   \begin{figure*}[h]
      \centering
      \includegraphics[width=17cm]{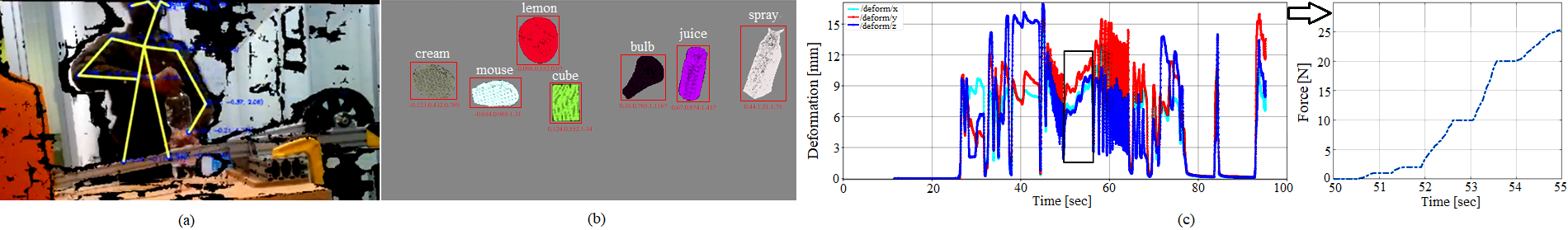}
      \caption{Visuo-tactile perception in formulating intuitive robotic tasks, (a) represents the aligned depth map on skeleton tracking of human subject performing desired actions on object in the scene, (b) is the point cloud of the scene with objects being recognized and their poses being enumerated locally, (c) is the 3D deformation output of sponge based tactile sensors attached to gripper and the mapped force profile of gripper using shallow neural network for interaction modulation.\\}
      \label{raw}
      \vspace{-15pt}
   \end{figure*}

   \begin{figure*}[h]
      \centering
      \includegraphics[width=17cm]{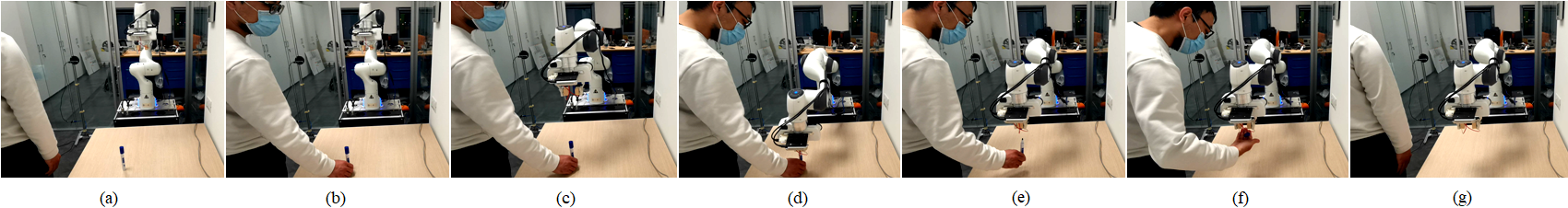}
      \caption{Robot collaborating with human on detaching marker from its cap, (a) represents the initial configuration of all the actors (human, robot and object), (b) illustrates the human griping marker from its bottom, in (c) the robot arm assumes pre-grasping posture based on the pose of active human arm wrist, (d) shows the robot arm grasping cap from its centroid posture, in (e) human pulls down the marker while the gripper holds the cap, in (f) the gripper releases cap on detecting open human palm, in (g) the robot arm returns to homing position following its human partner. (Video: \url{https://www.youtube.com/watch?v=j41rgnEavx4}).}
      \label{nine}
      \vspace{-15pt}
   \end{figure*}
   
The defined tasks are subsequently augmented with visual and tactile information for intuitive decision making. Hence, a standard skeleton tracking algorithm (i.e, deep CONVNETs) \cite{c11} is used for tracing the gestures of active human arm using a aligned depth map captured with RGBD tracking camera in Fig. \ref{raw} (a), which is registering the poses of 18 human joints in the local camera frame. Next, a modified RANSAC algorithm together with Support Vector Machine (SVM) classifier is used for object semantic segmentation and recognition respectively and the poses of candidate objects are enumerated by computing the centroid of their processed point cloud captured with RGBD detection camera in Fig. \ref{raw} (b). Moreover, for interaction tasks i.e, grasping and manipulation defined in SoT, the tactile feedback is explicitly being used. The installed tactile sensors \cite{c12} provide 3D deformation output and it is mapped to gripper's force profile using a shallow neural network (with 5 hidden neurons), in Fig. \ref{raw} (c). All the sensory observations are primarily in local reference frames and transformed into robot base using suitable transformations for homogeneous computations. 
   
\section{Results and Discussion}

To better evaluate the reliability and robustness of proposed intuitive task formulation, an industrial test scenario i,e, removing marker from the cap is considered, as shown in Fig. \ref{nine}. At first, all the characters i.e, robot, human and environment are in their initial configurations in Fig. \ref{nine} (a) and then the human tries to grip the marker from bottom in Fig. \ref{nine} (b) which is determined by tracking camera and thus the robot arm assumes the pre-grasping posture (i.e, 40 cm above the active human arm wrist) following human arm gesture in Fig. \ref{nine} (c). In this configuration, the detection camera recognizes the cap-marker pair in the scene and estimates their poses, which are used by the robot system to grasp the cap in Fig. \ref{nine} (d). Once the contact is established with cap-marker, the robot arm lifts it up (20 cm) in Fig. \ref{nine} (e) to provide a sufficient space to human partner to perform required action (pulling down) on the marker in Fig. \ref{nine} (f). After completing required task, the robot arm returns to its homing position following the human gesture in Fig. \ref{nine} (g).

\section{CONCLUSIONS}

This research proposed to exploit visuo-tactile information in formulating robotic tasks in accordance with human intuitions. Firstly, the visual feedback from tracking camera (Intel Realsense D435) using a aligned depth map estimated the gesture of active human arm to guide the robot arm to cooperate accordingly and later the detection camera (Intel Realsense D415) enumerated the object pose from the filtered point cloud, which was sent to gripper for respective grasping action. With the object being grasped, as detected by tactile sensors, the human executed designated task while the tactile sensors modulated the gripper's force profile to maintain the continuous desired contact with the object consistently.

\end{document}